\documentclass[aip,apl,reprint,preprintnumbers]{revtex4-1}
\usepackage{graphicx}
\usepackage{natbib}

\begin{document}

%\preprint{}

\title{THz-radiation production using dispersively-selected flat electron bunches} %Title of paper



\author{J. Thangaraj}
\email[]{jtobin@fnal.gov}

%\thanks{}
\affiliation{Accelerator Physics Center, Fermi National Accelerator Laboratory, Batavia, IL 60510, USA}

\author{P. Piot} 
\affiliation{Accelerator Physics Center, Fermi National Accelerator Laboratory, Batavia, IL 60510, USA}
\affiliation{Northern Illinois Center for Accelerator \& Detector Development and Department of Physics, Northern Illinois University, DeKalb IL 60115, USA} 

%\email[]{jtobin@fnal.gov}
%\thanks{}


\begin{abstract}
We propose an alternative scheme for a tunable THz radiation source generated by relativistic electron bunches. This technique relies on the combination of dispersive selection and flat electron bunch. The dispersive selection uses a slit mask inside a bunch compressor to transform the energy-chirped electron beam into a bunch train. The flat beam transformation boosts the frequency range of the THz source by reducing the beam emittance in one plane. This technique generates narrow-band THz radiation with a tuning range between 0.2 - 4 THz. Single frequency THz spectrum can also be generated by properly choosing the slit spacing, slit width, and the energy chirp.
 
\end{abstract}


\maketitle %\maketitle must follow title, authors, abstract and \pacs


  Accelerator-driven terahertz (THz) sources have attracted immense interest over a broad range of disciplines due to their ability to produce a high power, tunable radiation within compact footprint\cite{THzworkshop2013}. Accelerator-based THz sources combine a sub-picosecond relativistic electron bunch with an electromagnetic radiative process, e.g., the beam could either pass through a foil radiator to emit coherent transition radiation (CTR) or travel through a dipole to emit coherent synchrotron radiation (CSR)\cite{wuslac,carr2002high,flashctr}. The total spectral intensity of the emitted radiation from an electron bunch consisting of N electrons through a radiative electromagnetic process is given by\cite{saxon}:$\displaystyle\left(\frac{d^2I}{d\omega d\Omega}\right)_t=\left(\frac{d^2I}{d\omega d\Omega}\right)_e[N+N(N-1)|B_0(\omega)|^2]$, where $\omega= 2\pi f$ ($f$ is the frequency), $\displaystyle\left(\frac{d^2I}{d\omega d\Omega}\right)_e$ is the single electron spectral intensity and $B_0(\omega)= \sum\limits_{k=1}^{N} e^{i\omega t_k}$ is the bunching factor, where $t_k$ is the the longitudinal time coordinate of the $k^{th}$ electron inside the bunch. Other broad-band THz schemes include advanced acceleration schemes such as laser-driven plasma acceleration and ion-driven acceleration\cite{ionTHz,leemansTHz}. Narrow-band THz sources  uses a variety of techniques such as corrugated waveguide\cite{banecorrugated}, emittance exchanger\cite{piotTHz}, modulating the drive laser \cite{shenprl,Boscolonima}, echo-based\cite{echoTHz}, or dielectric based \cite{antipovTHz} schemes. In this letter, we propose a simple scheme for THz generation using a slit mask in an dispersive region of a linear accelerator to generate up to 4 THz using a 50 MeV beam. The achievable frequency range span 0.2 - 4 THz. All this scheme requires is a photoinjector and a bunch compressor both of which are a standard components at almost all modern and planned future linear accelerators.
    
   %We begin by describing the technique based on dispersive scraping to generate a comb of electron pulses for THz generation. In the next section, we introduce the idea of flat beam generation and show the advantage of beam phase-space manipulation to boost the THz frequency beyond the sub-THz regime ($<$ 1 THz) to multi-THz ($>$ 1 THz) regime. We also show the merits of this scheme versus others and ease of detection compared to other detection techniques.%

Magnetic bunch compressor are commonly incorporated in accelerators that drive free-electron lasers (FEL) to enhance the electron bunch peak current. Generating a train of sub-picosecond bunches using dispersive scraping in a chicane (four dipoles bending angle +,-,-,+) or in a dogleg (two dipoles separated by a drift) bunch compressor has been developed elsewhere\cite{Nguyennima,muggliatf}.  

\begin{figure}[h] % float placement: (h)ere, page (t)op, page (b)ottom, other (p)age
  \centering
  \includegraphics*[width=\columnwidth]{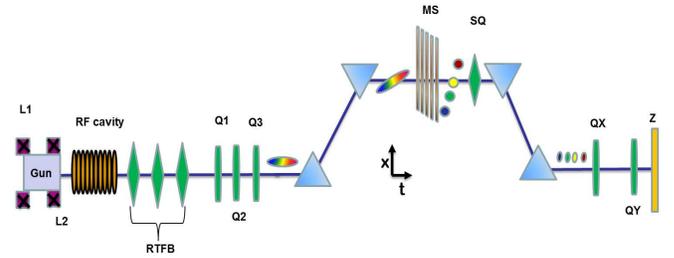}
  \caption{Schematic of the THz beamline: The RF photoinjector consist of a gun and two solenoid lenses (L1, L2). After existing the gun, the electron beam is acceleration off-crest in the RF cavity. This  energy-chirped beam is focussed using the quadrupoles (Q1, Q2, Q3) and then enters a chicane and is intercepted by a set of slit mask (MS) at the center. After the slit mask, some electrons are scattered while other pass through the chicane. At the end of the chicane transversely separated electron beam are transformed into longitudinally separated train of bunches. Blue (head) is higher energy and red (tail) is lower energy. The beam is focussed on the CTR aluminium foil (Z) using the quadrupole doublet (QX, QY)to extract the THz. The round to flat beam transformer (RTFB) section of the linac has three skew quadrupoles representd by diamond to generate a flat beam for multi-THz. There is another skewquad (SQ) close to the center of the chicane for diagnostics.  }
  \label{f1}
\end{figure}

Figure~{\ref{f1}} illustrate the principle of the proposed method. An electron beam is generated from a photoinjector and is then accelerated by a radio-frequency (RF) cavity. During acceleration, the electron beam gets an energy chirp - a time-dependent energy variation. The energy-chirped beam is then sent  through a straight section of the linac that includes quadrupole magnets (Q1, Q2, Q3 in Fig.~{\ref{f1}}) and then to the  bunch compressor. At the center of the bunch compressor, the bunch is intercepted  by  a slit mask (MS) which selectively scatters some of the electrons while other electrons are transmitted through the rest of the chicane. At the end of the chicane, such transversely separated beamlets are transformed  into a train of short bunches longitudinally. The spacing between the bunches and the length of each bunch is determined by several factors such as the dispersion of the chicane ($\eta$), the transverse betatron spot size of the beam at the mask, the width of the slit mask ($w$), the uncorrelated relative beam energy spread ($\sigma_u$) and the RF-energy chirp on the beam ($h$). The formula that relates the length of the bunch at the exit of chicane to the width of the slit is given by \cite{emmaspoiler}: $\sigma_z=\frac{1}{|\eta h|}{\sqrt{\eta^2\sigma_u^2+(1+hR_{56})^2[\Delta X^2+\varepsilon\beta]}}$, where $\sigma_z$ is the output bunch length, $R_{56}$ is the longitudinal dispersion of the chicane, $\Delta X=\frac{w}{2\sqrt3}$ is the rms width of the mask, $\varepsilon$ is the natural beam emittance, and $\beta$ is the betatron function at the mask. It can be seen that when $hR_{56}$ is large, the output bunch profile follows the mask profile ($\Delta X$). This can be done by making $|1+hR_{56}| >> 1 $. For a chicane in our convention $R_{56} < 0$ ($z > 0$ corresponds to the tail) , therefore by setting $h < 0$, the output bunch profile can be made to follow the mask profile. This technique is limited by the initial slice energy spread and emittance of the beam. We note that same function can also be reached by setting $h << \frac{-2}{R_{56}}$, which can become very large and impractical and in certain cases lead to overcompression. The above equation also indicates that to get a bunch train one should ensure that the betatron spot size  at the slit mask is less than the slit width ($\varepsilon\beta << (\Delta X)^2$). This can be done by properly setting the quadrupole magnet triplet (Q1, Q2, Q3) located upstream of the chicane to the right current setting.  In order to reveal the longitudinal structure, the skew quadrupole (SQ) can be powered on that couples the x-dispersion into the y-plane and therefore the vertical ($y$) axis on the screen downstream is transformed into a time axis\cite{emma-skewquad}. %Such a technique has been done in the context of studying CSR in a bunch compressor .%
Finally, we note that this scheme allows for pulse shaping other than a train of pulses: for e.g. a triangular wedge shaped collimator can be used to generate ramped bunches that have application in advanced accelerator-type applications.

Hence, in our scheme the magnetic chicane effectively acts to decompress the bunch.  By dispersing the beam inside the chicane, an $x-z$ correlation is introduced at the center of the chicane, where $x$ is the transverse position of the particle and $z$ its longitudinal position of the particle. Due to this high correlation, any variation in $x$ is then mapped onto $z$.  This scheme is different from \cite{emmaspoiler} where differential spoiling is used at high energy (few GeV) to generate femtosecond x-rays. Our scheme differs from it in two aspects: the low energy of our beam allows us to stop or scatter much of the beam using metallic slits and the bunch compressor is set to decompression. Also, our intrinsic relative energy spread is fairly high compared to that scheme because of the low energy of the beam. As mentioned above, our scheme differs from  \cite{muggliatf} by using a chicane instead of a dogleg and using the RF chirp as the tuning variable instead of using quadrupoles and an energy slit. We note that our scheme is more efficient since there is already an energy-chirp imparted naturally due to the longitudinal space-charge forces when the bunch exists the photoinjector that is favorable to our scheme (head is at high energy and tail is at lower energy) before it enters the RF cavity.

%\label{}

\begin{figure}[h] % float placement: (h)ere, page (t)op, page (b)ottom, other (p)age
  \centering
  \includegraphics*[width=\columnwidth]{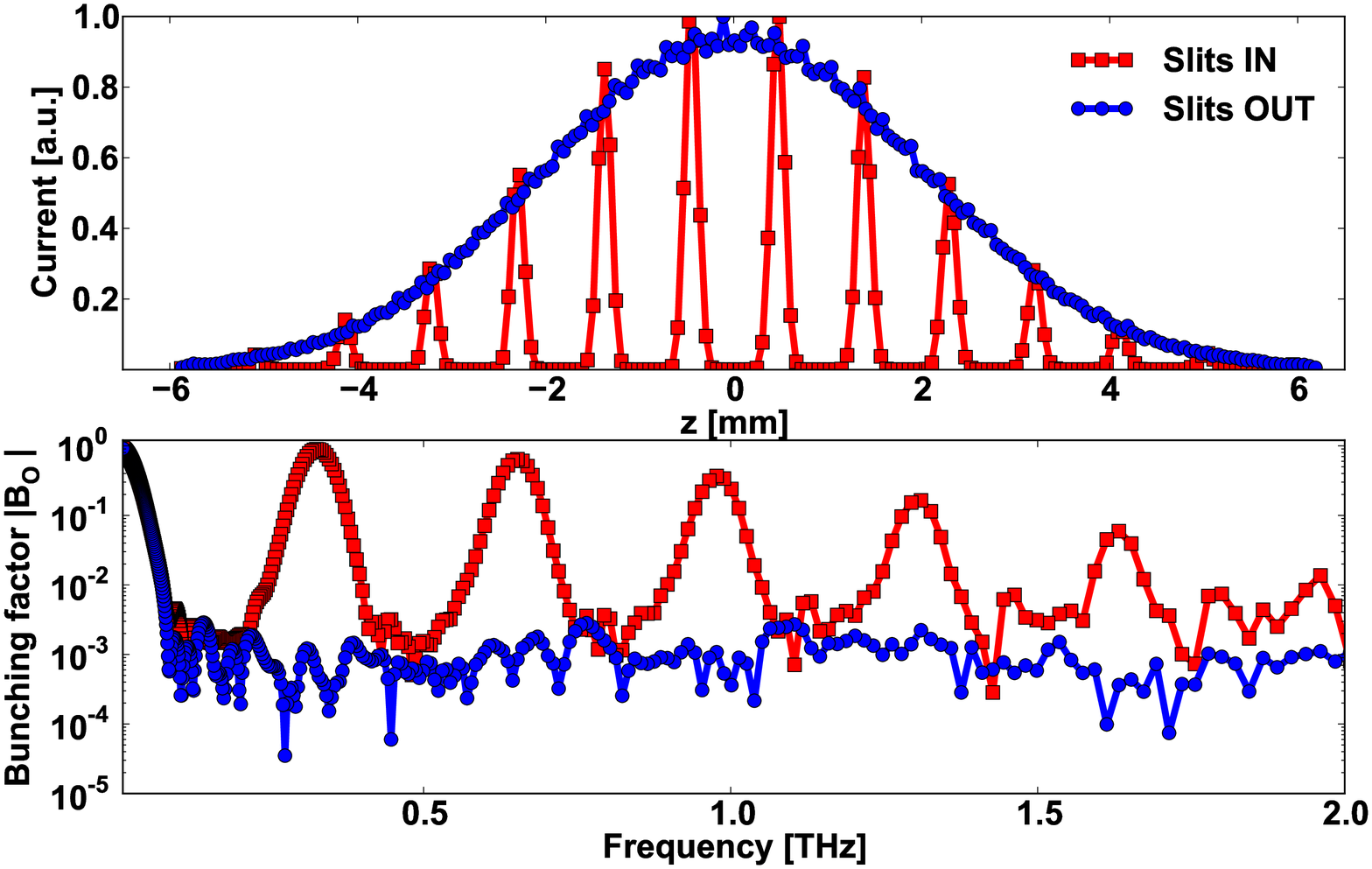}
  \caption{Normalized current profile of the electron bunch (top) and associated bunch form factor (bottom) with (red) and without (blue) the slits inserted. When the slits are inserted, the beam is bunched at sub-THz frequencies and hence the resonant enhancement in the frequency domain at harmonics of the bunching frequencies.}
  \label{f2}
\end{figure}

 We show through tracking simulation that our scheme can generate tunable, coherent sub-THz (i.e around or less than 1 THz) radiation. The particle tracking program ELEGANT\cite{elegant} was used for simulating the beam line. All the bending magnets are rectangular magnets. In all the simulation shown in this paper, CSR is taken into account.  Nominal values for slit width and slit spacing along with the beam and chicane parameters are shown in Table.~{\ref{t1}}. The initial phase-space distributions are assumed to be Gaussian. 
 %A quadrupole triplet in front of the chicane is used to do the $\beta$-function optimization for a minimum spot size at the slit mask. 
A linear energy chirp is assumed to be imparted by the RF-cavity. This is a fairly good approximation considering we are operating far from the off-crest with a decompressing phase. We note that in a laboratory beam the phase-space out of the photoinjector might still be distorted and further simulations are planned to understand such effects.

\begin{table}
\caption{\label{t1}Simulation parameters}
\begin{ruledtabular}
    \begin{tabular}{ l  l  l  }
    \hline
    \bf{Parameter} & \bf{Value} & \bf{Units} \\ \hline
    Initial emittance (x,y) & 0.5 & $\mu m$ \\ \hline
    Beam energy & 50 & MeV \\ \hline
    Initial slice energy spread & 5 & keV \\ \hline
    Initial bunch length & 0.8 & mm \\ \hline
    $\delta-z$ correlation (chirp) & [-10 ... -4] & 1/m  \\ \hline
    Charge & 100 & pC \\ \hline
    Slit spacing (center to center) & 1  & mm \\ \hline
    Slit width & 50 & $\mu m$       \\ \hline
    Number of particles & 10$^6$ & n/a. \\ \hline
    Dipole bending radius & 0.958 & m \\ \hline
    Dipole length & 0.301 & m \\ \hline
    Dipole angle & 18 & degrees \\ \hline
    $R_{56}$ & -18 & cm \\ \hline
    $\eta$ & -30 & cm \\ \hline
    \end{tabular}
\end{ruledtabular}
\end{table}

 Figure~{\ref{f2}} shows the current profile and the corresponding frequency spectrum from tracking simulation with and without the slits inside the beam line. When the slits are out, we get a single long, decompressed Gaussian bunch and the frequency spectrum obtained does not extend into the THz frequencies and is limited by the long bunch length.  However when the slits are inserted, we obtain a train of short bunches and the frequency spectrum has a fundamental and its harmonics with a narrow bandwidth. The relationship between the number of bunches in a train, the period of the bunch train, the rms width of the bunch and the frequency spectrum is given in \cite{piotTHz}. By tuning the RF-chirp on the electron beam prior to the chicane, the fundamental THz frequency can be tuned.  The upper limit of the THz frequency is limited by the uncorrelated relative energy spread and the normalized emittance of the beam.

\begin{figure}[h] % float placement: (h)ere, page (t)op, page (b)ottom, other (p)age
  \centering
    % Use this to determine the width of the figure.
  \includegraphics[width=\columnwidth]{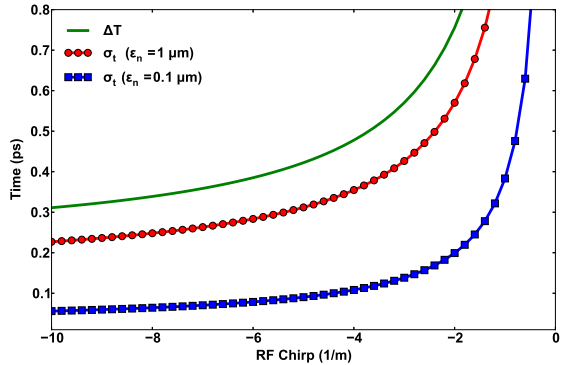}
  \caption{Effect of emittance on bunch train formation. Microbunch period $\Delta T$ and rms duration $\sigma_t$ as a function of RF chirp. For $\sigma_t$ two cases of emittance $\varepsilon_n=1 \ \mu m$ and $0.1 \ \mu m$ are considered. Above the solid-circled line region ($\varepsilon_n=1 \ \mu m$), the $\Delta T$ (solid line) is close to $\sigma_t$ and thus smears train formation but in the region above the solid-square line region ($\varepsilon_n=0.1 \ \mu m$), the lower emittance resolves the individual bunches because $\Delta T > 4\sigma_t$. Slit-width ($\Delta X = 50 \ \mu m$), slit spacing ($D = 100 \ \mu m$) and $\beta = 0.5 \ m$. }
  \label{f3}
\end{figure}

\begin{figure}[h] % float placement: (h)ere, page (t)op, page (b)ottom, other (p)age
  \centering
  \includegraphics*[width=\columnwidth]{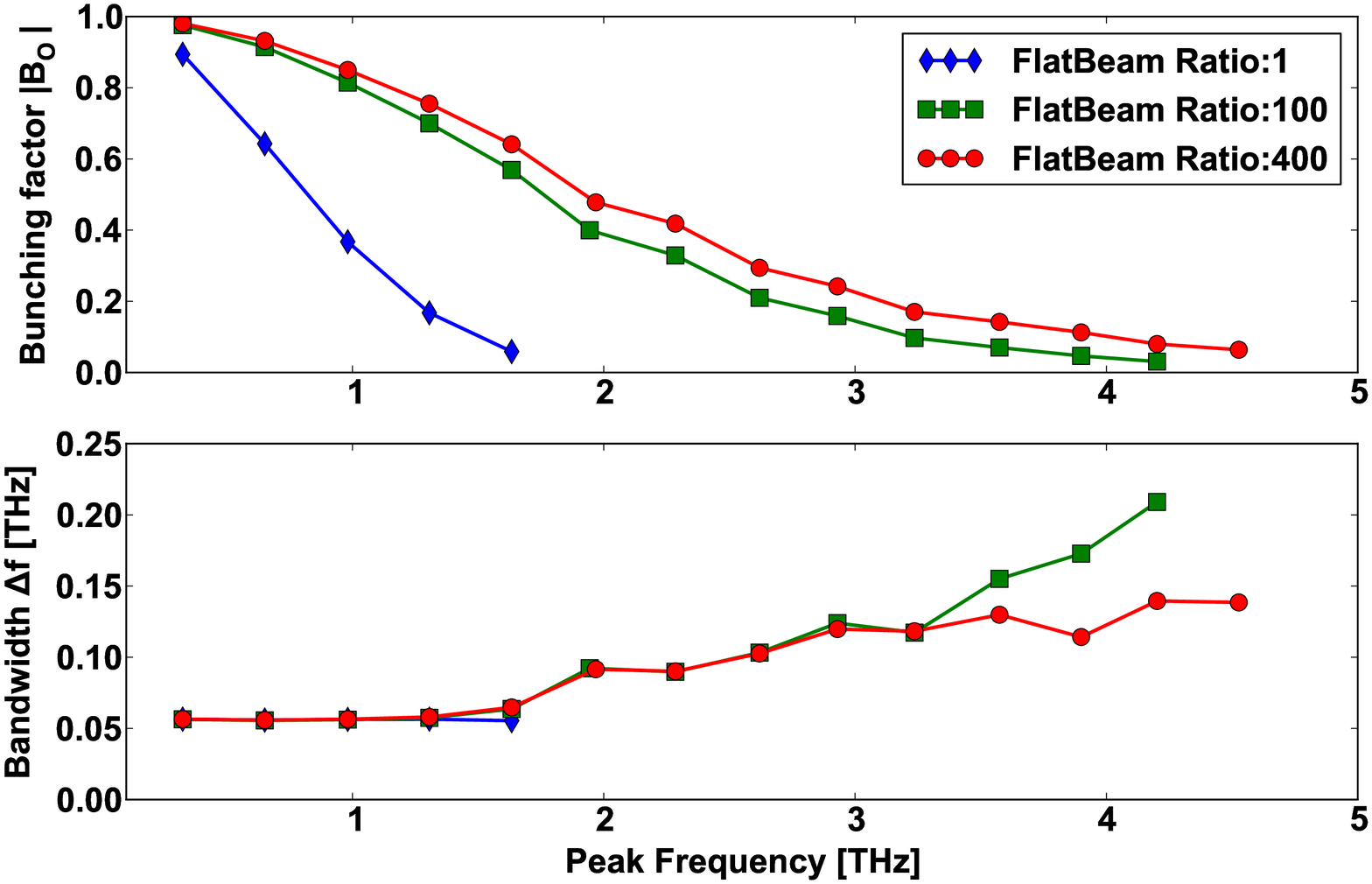}
  \caption{Boosted THz spectrum due to the flat-beam transformation showing the bunch form factor (top) and the bandwidth (bottom) extending well above 1 THz upto 4 THz compared with no flat-beam generation.}
  \label{f4}
\end{figure}

While the slit-based technique is capable of  generating sub-THz frequencies, it is non-trivial to go above 1 THz without additional complexity. In order to go above the THz barrier, one needs smaller mask width but then the emittance requirement becomes challenging ($\varepsilon\beta << (\Delta X)^2$). Figure~{\ref{f3}} illustrates the effect of the normalized emittance on the formation of the bunch train for a given slit spacing. In order to get a bunch train, the spacing between the bunches ($\Delta T$) must be larger than the bunch duration of the individual bunches ($\sigma_t=\frac{\sigma_z}{c}$) (typically, $\Delta T > 4\sigma_t$). The microbunch period is $\Delta T=\frac{D}{\eta |h C| c}$, where $D$ is the slits spacing, $C$ is the compression factor given by $C=(1+hR_{56})^{-1}$ and $c$ is the speed of light. As shown in Fig.~{\ref{f3}}, lower emittance beam allows bunch train formation by producing shorter individual bunches for a fixed $\Delta T$. One way to achieve low emittance would be to operate the linac at a lower charge (10 pC) but when going through the slits most of charge (upto 90\%) could be lost. Another way to achieve low emittance in one plane only for e.g. in the horizontal plane is through flat-beam transformation. In order to generate a flat beam, the photocathode is immersed in an axial magnetic field which generates a magnetized electron beam. After acceleration, a set of three skew quadrupoles (RFBT in Fig.~{\ref{f1}}), is used to transform the magnetized beam into a flat beam. Such flat-beam transformation have been studied theoretically and demonstrated experimentally \cite{brinkmanflat,piotflat}. A flat beam ratio of $\varepsilon_x:\varepsilon_y$ of 100  has been experimentally demonstrated at low energies using the Fermilab A0 photoinjector.  Note the product of the emittances $\varepsilon_x\varepsilon_y$ remains constant before and after the flat-beam transformation. Therefore, to achieve the required boost in the THz frequency and break the sub-THz barrier, we use flat-beam transformation in the linac. In order to demonstrate this, we use ELEGANT simulation. We use an emittance ratio of 100 and 400 which is consistent with simulation\cite{astaflat}.  The results shown in Fig.~{\ref{f4}} indicates that the use of flat beam transformations helps to generate higher THz frequencies for a given slit spacing and width. The flat-beam transformation not only extends the maximum THz frequency but also improves the bunch form factor at lower frequencies as well. A scan over various emittance ratio and RF-chirp shows that frequencies as high as 4 THz can be obtained.  In a superconducting linac, the RF-chirp can be controlled in a very precise manner with longitudinal feedback systems. Thus combining flat-beam technique, which can be done in any modern photoinjector linac using appropriate skew quadrupole magnets and a chicane equipped with a transverse mask, we can generate tunable multi-THz frequencies. In order to extract the THz radiation outside the beam pipe, we use a quadrupole doublet (QX, QY) followed by a CTR aluminium foil (Z shown in Fig.~{\ref{f1}). Our simulation shows that a rms (round) spot size  of $\sigma_r$=0.2 mm on both planes can be obtained at the screen using the doublet. This implies an upper cut-off frequency due to the transverse spot size of $f_u \sim \frac{\gamma c}{2\pi\sigma_r}$ of 23 THz which is well above our highest frequency of our scheme ($\gamma$=100 at 50 MeV ).  %We point out that during the experiment in order to tune the linac for the beam bunching, the skew quadrupole (SQ) can be used. When powered,the skew quadrupole couples the x-dispersion into the y-plane and therefore the y-axis on the screen (Z) is transformed into a time axis. Successful experimental studies have been done using such a technique in the context of studying CSR in a bunch compressor \cite{emma-skewquad}.

\begin{figure}[h] % float placement: (h)ere, page (t)op, page (b)ottom, other (p)age
  \centering
%\showthe\columnwidth
  \includegraphics*[width=\columnwidth]{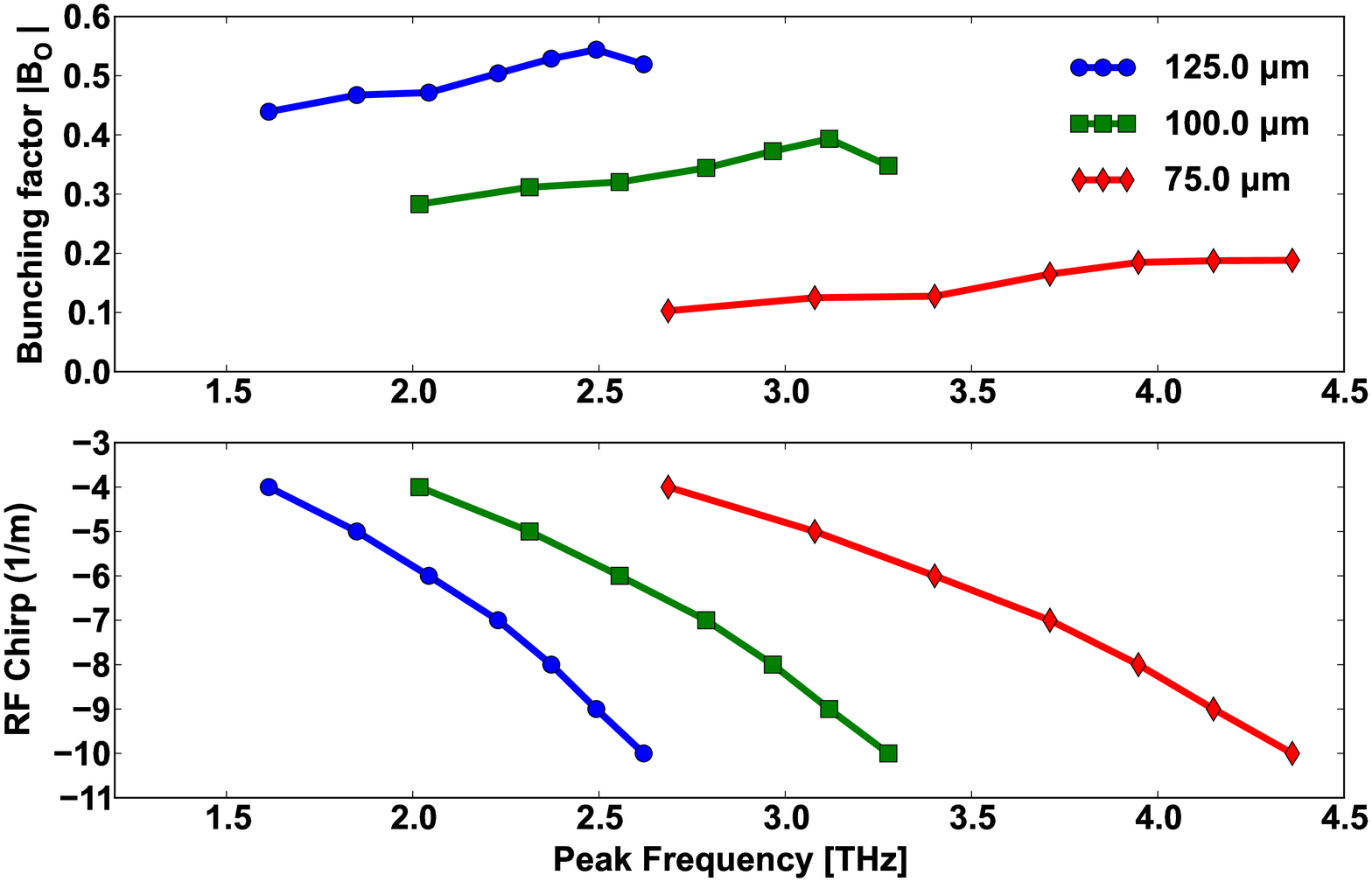}
  \caption{The bunching factor (above) of the single spike THz spectrum along with the required RF-chirp (below) as a function of the spacing of the slits. By picking a specific slit spacing and appropriate RF-chirp, a narrow-band single frequency THz spectrum can be generated.}
  \label{f5}
\end{figure}

While both the fundamental frequency and its harmonics are present in the bunch due to the flat-beam transformation, sometimes only a single THz frequency might be preferred by  users. This can be done by choosing the appropriate slit spacing and the width and supplying the correct RF-chirp. Figure~{\ref{f5}} shows the effect of varying the slit-spacing (D) by choosing smaller-width slits (20 $\mu m$) and RF-chirp. For this simulation, all other parameters remaining constant (Table.~{\ref{t1}}), a flat beam ratio of $\varepsilon_x:\varepsilon_y$=1:400 was used \cite{astaflat}. Proper choice of slit-spacing and RF-chirp allows a tunable range of 1-4 THz with a single frequency THz spectrum. A movable plate mounted with slits of different width and different spacing can easily be accommodated in a stepper motor controlled actuator to add this useful feature to the machine.

In summary, we have proposed and investigated via computer simulations a THz generation scheme that combines dispersive selection with flat electron beams. The advantage of this technique is its simplicity, tunability and low cost. The scheme does not require any additional hardware such as lasers, undulator, transverse deflecting cavity.  Our scheme can be readily deployed in any linac that uses low energy compression such as ASTA \cite{asta}, FLUTE\cite{flute}. By using low emittance beam  via flat-beam transformation in only one plane, tunable THz source covering 0.2 - 4 THz can be achieved. This scheme is also scalable to any superconducting linac as the only requirement is that the slit material should be able to withstand the heat load due to the multi-pulse structure of the electron bunch. Currently, experiments are planned at Fermilab's ASTA facility using this scheme and we anticipate this technique to be useful for other accelerators.

% If in two-column mode, this environment will change to single-column format so that long equations can be displayed. 
%\begin{widetext}
%$$\mbox{put long equation here}$$
%\end{widetext}

% Figures should be put into the text as floats. 
% Use the graphics or graphicx packages (distributed with LaTeX2e).
% Put the label that you will use with \ref{} command in the braces of the \label{} command.
%
% \begin{figure}
% \end{figure}

% \begin{table}
% \end{tabular}
% \end{table}


We would like to thank M. Borland for his support in ELEGANT simulation. One of us (J. T.) would like to thank Randy-Thurman Keup for clarifying issues on THz detection. The work was supported by the Fermi Research Alliance, LLC under the U.S. Department of Energy.

\bibliographystyle{apsrev4-1}
\bibliography{zslicer}

\begin{thebibliography}{23}%
\makeatletter
\providecommand \@ifxundefined [1]{%
 \@ifx{#1\undefined}
}%
\providecommand \@ifnum [1]{%
 \ifnum #1\expandafter \@firstoftwo
 \else \expandafter \@secondoftwo
 \fi
}%
\providecommand \@ifx [1]{%
 \ifx #1\expandafter \@firstoftwo
 \else \expandafter \@secondoftwo
 \fi
}%
\providecommand \natexlab [1]{#1}%
\providecommand \enquote  [1]{``#1''}%
\providecommand \bibnamefont  [1]{#1}%
\providecommand \bibfnamefont [1]{#1}%
\providecommand \citenamefont [1]{#1}%
\providecommand \href@noop [0]{\@secondoftwo}%
\providecommand \href [0]{\begingroup \@sanitize@url \@href}%
\providecommand \@href[1]{\@@startlink{#1}\@@href}%
\providecommand \@@href[1]{\endgroup#1\@@endlink}%
\providecommand \@sanitize@url [0]{\catcode `\\12\catcode `\$12\catcode
  `\&12\catcode `\#12\catcode `\^12\catcode `\_12\catcode `\%12\relax}%
\providecommand \@@startlink[1]{}%
\providecommand \@@endlink[0]{}%
\providecommand \url  [0]{\begingroup\@sanitize@url \@url }%
\providecommand \@url [1]{\endgroup\@href {#1}{\urlprefix }}%
\providecommand \urlprefix  [0]{URL }%
\providecommand \Eprint [0]{\href }%
\providecommand \doibase [0]{http://dx.doi.org/}%
\providecommand \selectlanguage [0]{\@gobble}%
\providecommand \bibinfo  [0]{\@secondoftwo}%
\providecommand \bibfield  [0]{\@secondoftwo}%
\providecommand \translation [1]{[#1]}%
\providecommand \BibitemOpen [0]{}%
\providecommand \bibitemStop [0]{}%
\providecommand \bibitemNoStop [0]{.\EOS\space}%
\providecommand \EOS [0]{\spacefactor3000\relax}%
\providecommand \BibitemShut  [1]{\csname bibitem#1\endcsname}%
\let\auto@bib@innerbib\@empty
%</preamble>
\bibitem [{\citenamefont {Wen}\ \emph {et~al.}(2013)\citenamefont {Wen},
  \citenamefont {Kim}, \citenamefont {Zholents}, \citenamefont {Byrd},\ and\
  \citenamefont {Cavalleri}}]{THzworkshop2013}%
  \BibitemOpen
  \bibfield  {author} {\bibinfo {author} {\bibfnamefont {H.}~\bibnamefont
  {Wen}}, \bibinfo {author} {\bibfnamefont {K.-J.}\ \bibnamefont {Kim}},
  \bibinfo {author} {\bibfnamefont {A.}~\bibnamefont {Zholents}}, \bibinfo
  {author} {\bibfnamefont {J.}~\bibnamefont {Byrd}}, \ and\ \bibinfo {author}
  {\bibfnamefont {A.}~\bibnamefont {Cavalleri}},\ }\href {\doibase
  10.1063/1.4790426} {\bibfield  {journal} {\bibinfo  {journal} {Review of
  Scientific Instruments}\ }\textbf {\bibinfo {volume} {84}},\ \bibinfo {eid}
  {022501} (\bibinfo {year} {2013})}\BibitemShut {NoStop}%
\bibitem [{\citenamefont {Wu}\ \emph {et~al.}(2013)\citenamefont {Wu},
  \citenamefont {Fisher}, \citenamefont {Goodfellow}, \citenamefont {Fuchs},
  \citenamefont {Daranciang}, \citenamefont {Hogan}, \citenamefont {Loos},\
  and\ \citenamefont {Lindenberg}}]{wuslac}%
  \BibitemOpen
  \bibfield  {author} {\bibinfo {author} {\bibfnamefont {Z.}~\bibnamefont
  {Wu}}, \bibinfo {author} {\bibfnamefont {A.~S.}\ \bibnamefont {Fisher}},
  \bibinfo {author} {\bibfnamefont {J.}~\bibnamefont {Goodfellow}}, \bibinfo
  {author} {\bibfnamefont {M.}~\bibnamefont {Fuchs}}, \bibinfo {author}
  {\bibfnamefont {D.}~\bibnamefont {Daranciang}}, \bibinfo {author}
  {\bibfnamefont {M.}~\bibnamefont {Hogan}}, \bibinfo {author} {\bibfnamefont
  {H.}~\bibnamefont {Loos}}, \ and\ \bibinfo {author} {\bibfnamefont
  {A.}~\bibnamefont {Lindenberg}},\ }\href {\doibase 10.1063/1.4790427}
  {\bibfield  {journal} {\bibinfo  {journal} {Review of Scientific
  Instruments}\ }\textbf {\bibinfo {volume} {84}},\ \bibinfo {eid} {022701}
  (\bibinfo {year} {2013})}\BibitemShut {NoStop}%
\bibitem [{\citenamefont {Carr}\ \emph {et~al.}(2002)\citenamefont {Carr},
  \citenamefont {Martin}, \citenamefont {McKinney}, \citenamefont {Jordan},
  \citenamefont {Neil},\ and\ \citenamefont {Williams}}]{carr2002high}%
  \BibitemOpen
  \bibfield  {author} {\bibinfo {author} {\bibfnamefont {G.}~\bibnamefont
  {Carr}}, \bibinfo {author} {\bibfnamefont {M.}~\bibnamefont {Martin}},
  \bibinfo {author} {\bibfnamefont {W.}~\bibnamefont {McKinney}}, \bibinfo
  {author} {\bibfnamefont {K.}~\bibnamefont {Jordan}}, \bibinfo {author}
  {\bibfnamefont {G.}~\bibnamefont {Neil}}, \ and\ \bibinfo {author}
  {\bibfnamefont {G.}~\bibnamefont {Williams}},\ }\href@noop {} {\bibfield
  {journal} {\bibinfo  {journal} {Nature}\ }\textbf {\bibinfo {volume} {420}},\
  \bibinfo {pages} {153} (\bibinfo {year} {2002})}\BibitemShut {NoStop}%
\bibitem [{\citenamefont {Casalbuoni}\ \emph {et~al.}(2009)\citenamefont
  {Casalbuoni}, \citenamefont {Schmidt}, \citenamefont {Schm\"user},
  \citenamefont {Arsov},\ and\ \citenamefont {Wesch}}]{flashctr}%
  \BibitemOpen
  \bibfield  {author} {\bibinfo {author} {\bibfnamefont {S.}~\bibnamefont
  {Casalbuoni}}, \bibinfo {author} {\bibfnamefont {B.}~\bibnamefont {Schmidt}},
  \bibinfo {author} {\bibfnamefont {P.}~\bibnamefont {Schm\"user}}, \bibinfo
  {author} {\bibfnamefont {V.}~\bibnamefont {Arsov}}, \ and\ \bibinfo {author}
  {\bibfnamefont {S.}~\bibnamefont {Wesch}},\ }\href {\doibase
  10.1103/PhysRevSTAB.12.030705} {\bibfield  {journal} {\bibinfo  {journal}
  {Phys. Rev. ST Accel. Beams}\ }\textbf {\bibinfo {volume} {12}},\ \bibinfo
  {pages} {030705} (\bibinfo {year} {2009})}\BibitemShut {NoStop}%
\bibitem [{\citenamefont {Nodvick}\ and\ \citenamefont {Saxon}(1954)}]{saxon}%
  \BibitemOpen
  \bibfield  {author} {\bibinfo {author} {\bibfnamefont {J.~S.}\ \bibnamefont
  {Nodvick}}\ and\ \bibinfo {author} {\bibfnamefont {D.~S.}\ \bibnamefont
  {Saxon}},\ }\href {\doibase 10.1103/PhysRev.96.180} {\bibfield  {journal}
  {\bibinfo  {journal} {Phys. Rev.}\ }\textbf {\bibinfo {volume} {96}},\
  \bibinfo {pages} {180} (\bibinfo {year} {1954})}\BibitemShut {NoStop}%
\bibitem [{\citenamefont {Gopal}\ \emph {et~al.}(2013)\citenamefont {Gopal},
  \citenamefont {Herzer}, \citenamefont {Schmidt}, \citenamefont {Singh},
  \citenamefont {Reinhard}, \citenamefont {Ziegler}, \citenamefont {Br\"ommel},
  \citenamefont {Karmakar}, \citenamefont {Gibbon}, \citenamefont {Dillner},
  \citenamefont {May}, \citenamefont {Meyer},\ and\ \citenamefont
  {Paulus}}]{ionTHz}%
  \BibitemOpen
  \bibfield  {author} {\bibinfo {author} {\bibfnamefont {A.}~\bibnamefont
  {Gopal}}, \bibinfo {author} {\bibfnamefont {S.}~\bibnamefont {Herzer}},
  \bibinfo {author} {\bibfnamefont {A.}~\bibnamefont {Schmidt}}, \bibinfo
  {author} {\bibfnamefont {P.}~\bibnamefont {Singh}}, \bibinfo {author}
  {\bibfnamefont {A.}~\bibnamefont {Reinhard}}, \bibinfo {author}
  {\bibfnamefont {W.}~\bibnamefont {Ziegler}}, \bibinfo {author} {\bibfnamefont
  {D.}~\bibnamefont {Br\"ommel}}, \bibinfo {author} {\bibfnamefont
  {A.}~\bibnamefont {Karmakar}}, \bibinfo {author} {\bibfnamefont
  {P.}~\bibnamefont {Gibbon}}, \bibinfo {author} {\bibfnamefont
  {U.}~\bibnamefont {Dillner}}, \bibinfo {author} {\bibfnamefont
  {T.}~\bibnamefont {May}}, \bibinfo {author} {\bibfnamefont {H.-G.}\
  \bibnamefont {Meyer}}, \ and\ \bibinfo {author} {\bibfnamefont {G.~G.}\
  \bibnamefont {Paulus}},\ }\href {\doibase 10.1103/PhysRevLett.111.074802}
  {\bibfield  {journal} {\bibinfo  {journal} {Phys. Rev. Lett.}\ }\textbf
  {\bibinfo {volume} {111}},\ \bibinfo {pages} {074802} (\bibinfo {year}
  {2013})}\BibitemShut {NoStop}%
\bibitem [{\citenamefont {Leemans}\ \emph {et~al.}(2003)\citenamefont
  {Leemans}, \citenamefont {Geddes}, \citenamefont {Faure}, \citenamefont
  {T\'oth}, \citenamefont {van Tilborg}, \citenamefont {Schroeder},
  \citenamefont {Esarey}, \citenamefont {Fubiani}, \citenamefont {Auerbach},
  \citenamefont {Marcelis}, \citenamefont {Carnahan}, \citenamefont {Kaindl},
  \citenamefont {Byrd},\ and\ \citenamefont {Martin}}]{leemansTHz}%
  \BibitemOpen
  \bibfield  {author} {\bibinfo {author} {\bibfnamefont {W.~P.}\ \bibnamefont
  {Leemans}}, \bibinfo {author} {\bibfnamefont {C.~G.~R.}\ \bibnamefont
  {Geddes}}, \bibinfo {author} {\bibfnamefont {J.}~\bibnamefont {Faure}},
  \bibinfo {author} {\bibfnamefont {C.}~\bibnamefont {T\'oth}}, \bibinfo
  {author} {\bibfnamefont {J.}~\bibnamefont {van Tilborg}}, \bibinfo {author}
  {\bibfnamefont {C.~B.}\ \bibnamefont {Schroeder}}, \bibinfo {author}
  {\bibfnamefont {E.}~\bibnamefont {Esarey}}, \bibinfo {author} {\bibfnamefont
  {G.}~\bibnamefont {Fubiani}}, \bibinfo {author} {\bibfnamefont
  {D.}~\bibnamefont {Auerbach}}, \bibinfo {author} {\bibfnamefont
  {B.}~\bibnamefont {Marcelis}}, \bibinfo {author} {\bibfnamefont {M.~A.}\
  \bibnamefont {Carnahan}}, \bibinfo {author} {\bibfnamefont {R.~A.}\
  \bibnamefont {Kaindl}}, \bibinfo {author} {\bibfnamefont {J.}~\bibnamefont
  {Byrd}}, \ and\ \bibinfo {author} {\bibfnamefont {M.~C.}\ \bibnamefont
  {Martin}},\ }\href {\doibase 10.1103/PhysRevLett.91.074802} {\bibfield
  {journal} {\bibinfo  {journal} {Phys. Rev. Lett.}\ }\textbf {\bibinfo
  {volume} {91}},\ \bibinfo {pages} {074802} (\bibinfo {year}
  {2003})}\BibitemShut {NoStop}%
\bibitem [{\citenamefont {Bane}\ and\ \citenamefont
  {Stupakov}(2012)}]{banecorrugated}%
  \BibitemOpen
  \bibfield  {author} {\bibinfo {author} {\bibfnamefont {K.}~\bibnamefont
  {Bane}}\ and\ \bibinfo {author} {\bibfnamefont {G.}~\bibnamefont
  {Stupakov}},\ }\href {\doibase http://dx.doi.org/10.1016/j.nima.2012.02.028}
  {\bibfield  {journal} {\bibinfo  {journal} {Nucl. Instrum. Methods. Phys.
  Res. A}\ }\textbf {\bibinfo {volume} {677}},\ \bibinfo {pages} {67 }
  (\bibinfo {year} {2012})}\BibitemShut {NoStop}%
\bibitem [{\citenamefont {Piot}\ \emph {et~al.}(2011)\citenamefont {Piot},
  \citenamefont {Sun}, \citenamefont {Maxwell}, \citenamefont {Ruan},
  \citenamefont {Lumpkin}, \citenamefont {Rihaoui},\ and\ \citenamefont
  {Thurman-Keup}}]{piotTHz}%
  \BibitemOpen
  \bibfield  {author} {\bibinfo {author} {\bibfnamefont {P.}~\bibnamefont
  {Piot}}, \bibinfo {author} {\bibfnamefont {Y.-E.}\ \bibnamefont {Sun}},
  \bibinfo {author} {\bibfnamefont {T.~J.}\ \bibnamefont {Maxwell}}, \bibinfo
  {author} {\bibfnamefont {J.}~\bibnamefont {Ruan}}, \bibinfo {author}
  {\bibfnamefont {A.~H.}\ \bibnamefont {Lumpkin}}, \bibinfo {author}
  {\bibfnamefont {M.~M.}\ \bibnamefont {Rihaoui}}, \ and\ \bibinfo {author}
  {\bibfnamefont {R.}~\bibnamefont {Thurman-Keup}},\ }\href {\doibase
  10.1063/1.3604017} {\bibfield  {journal} {\bibinfo  {journal} {Applied
  Physics Letters}\ }\textbf {\bibinfo {volume} {98}},\ \bibinfo {eid} {261501}
  (\bibinfo {year} {2011})}\BibitemShut {NoStop}%
\bibitem [{\citenamefont {Shen}\ \emph {et~al.}(2011)\citenamefont {Shen},
  \citenamefont {Yang}, \citenamefont {Carr}, \citenamefont {Hidaka},
  \citenamefont {Murphy},\ and\ \citenamefont {Wang}}]{shenprl}%
  \BibitemOpen
  \bibfield  {author} {\bibinfo {author} {\bibfnamefont {Y.}~\bibnamefont
  {Shen}}, \bibinfo {author} {\bibfnamefont {X.}~\bibnamefont {Yang}}, \bibinfo
  {author} {\bibfnamefont {G.~L.}\ \bibnamefont {Carr}}, \bibinfo {author}
  {\bibfnamefont {Y.}~\bibnamefont {Hidaka}}, \bibinfo {author} {\bibfnamefont
  {J.~B.}\ \bibnamefont {Murphy}}, \ and\ \bibinfo {author} {\bibfnamefont
  {X.}~\bibnamefont {Wang}},\ }\href {\doibase 10.1103/PhysRevLett.107.204801}
  {\bibfield  {journal} {\bibinfo  {journal} {Phys. Rev. Lett.}\ }\textbf
  {\bibinfo {volume} {107}},\ \bibinfo {pages} {204801} (\bibinfo {year}
  {2011})}\BibitemShut {NoStop}%
\bibitem [{\citenamefont {Boscolo}\ \emph {et~al.}(2007)\citenamefont
  {Boscolo}, \citenamefont {Ferrario}, \citenamefont {Boscolo}, \citenamefont
  {Castelli},\ and\ \citenamefont {Cialdi}}]{Boscolonima}%
  \BibitemOpen
  \bibfield  {author} {\bibinfo {author} {\bibfnamefont {M.}~\bibnamefont
  {Boscolo}}, \bibinfo {author} {\bibfnamefont {M.}~\bibnamefont {Ferrario}},
  \bibinfo {author} {\bibfnamefont {I.}~\bibnamefont {Boscolo}}, \bibinfo
  {author} {\bibfnamefont {F.}~\bibnamefont {Castelli}}, \ and\ \bibinfo
  {author} {\bibfnamefont {S.}~\bibnamefont {Cialdi}},\ }\href {\doibase
  http://dx.doi.org/10.1016/j.nima.2007.04.129} {\bibfield  {journal} {\bibinfo
   {journal} {Nucl. Instrum. Methods. Phys. Res. A}\ }\textbf {\bibinfo
  {volume} {577}},\ \bibinfo {pages} {409 } (\bibinfo {year}
  {2007})}\BibitemShut {NoStop}%
\bibitem [{\citenamefont {Dunning}\ \emph {et~al.}(2012)\citenamefont
  {Dunning}, \citenamefont {Hast}, \citenamefont {Hemsing}, \citenamefont
  {Jobe}, \citenamefont {McCormick}, \citenamefont {Nelson}, \citenamefont
  {Raubenheimer}, \citenamefont {Soong}, \citenamefont {Szalata}, \citenamefont
  {Walz}, \citenamefont {Weathersby},\ and\ \citenamefont {Xiang}}]{echoTHz}%
  \BibitemOpen
  \bibfield  {author} {\bibinfo {author} {\bibfnamefont {M.}~\bibnamefont
  {Dunning}}, \bibinfo {author} {\bibfnamefont {C.}~\bibnamefont {Hast}},
  \bibinfo {author} {\bibfnamefont {E.}~\bibnamefont {Hemsing}}, \bibinfo
  {author} {\bibfnamefont {K.}~\bibnamefont {Jobe}}, \bibinfo {author}
  {\bibfnamefont {D.}~\bibnamefont {McCormick}}, \bibinfo {author}
  {\bibfnamefont {J.}~\bibnamefont {Nelson}}, \bibinfo {author} {\bibfnamefont
  {T.~O.}\ \bibnamefont {Raubenheimer}}, \bibinfo {author} {\bibfnamefont
  {K.}~\bibnamefont {Soong}}, \bibinfo {author} {\bibfnamefont
  {Z.}~\bibnamefont {Szalata}}, \bibinfo {author} {\bibfnamefont
  {D.}~\bibnamefont {Walz}}, \bibinfo {author} {\bibfnamefont {S.}~\bibnamefont
  {Weathersby}}, \ and\ \bibinfo {author} {\bibfnamefont {D.}~\bibnamefont
  {Xiang}},\ }\href {\doibase 10.1103/PhysRevLett.109.074801} {\bibfield
  {journal} {\bibinfo  {journal} {Phys. Rev. Lett.}\ }\textbf {\bibinfo
  {volume} {109}},\ \bibinfo {pages} {074801} (\bibinfo {year}
  {2012})}\BibitemShut {NoStop}%
\bibitem [{\citenamefont {Antipov}\ \emph {et~al.}(2013)\citenamefont
  {Antipov}, \citenamefont {Jing}, \citenamefont {Schoessow}, \citenamefont
  {Kanareykin}, \citenamefont {Yakimenko}, \citenamefont {Zholents},\ and\
  \citenamefont {Gai}}]{antipovTHz}%
  \BibitemOpen
  \bibfield  {author} {\bibinfo {author} {\bibfnamefont {S.}~\bibnamefont
  {Antipov}}, \bibinfo {author} {\bibfnamefont {C.}~\bibnamefont {Jing}},
  \bibinfo {author} {\bibfnamefont {P.}~\bibnamefont {Schoessow}}, \bibinfo
  {author} {\bibfnamefont {A.}~\bibnamefont {Kanareykin}}, \bibinfo {author}
  {\bibfnamefont {V.}~\bibnamefont {Yakimenko}}, \bibinfo {author}
  {\bibfnamefont {A.}~\bibnamefont {Zholents}}, \ and\ \bibinfo {author}
  {\bibfnamefont {W.}~\bibnamefont {Gai}},\ }\href {\doibase 10.1063/1.4790432}
  {\bibfield  {journal} {\bibinfo  {journal} {Review of Scientific
  Instruments}\ }\textbf {\bibinfo {volume} {84}},\ \bibinfo {eid} {022706}
  (\bibinfo {year} {2013})}\BibitemShut {NoStop}%
\bibitem [{\citenamefont {Nguyen}\ and\ \citenamefont
  {Carlsten}(1996)}]{Nguyennima}%
  \BibitemOpen
  \bibfield  {author} {\bibinfo {author} {\bibfnamefont {D.}~\bibnamefont
  {Nguyen}}\ and\ \bibinfo {author} {\bibfnamefont {B.}~\bibnamefont
  {Carlsten}},\ }\href {\doibase
  http://dx.doi.org/10.1016/0168-9002(95)01372-5} {\bibfield  {journal}
  {\bibinfo  {journal} {Nucl. Instrum. Methods. Phys. Res. A}\ }\textbf
  {\bibinfo {volume} {375}},\ \bibinfo {pages} {597 } (\bibinfo {year}
  {1996})},\ \bibinfo {note} {proceedings of the 17th International Free
  Electron Laser Conference}\BibitemShut {NoStop}%
\bibitem [{\citenamefont {Muggli}\ \emph {et~al.}(2008)\citenamefont {Muggli},
  \citenamefont {Yakimenko}, \citenamefont {Babzien}, \citenamefont {Kallos},\
  and\ \citenamefont {Kusche}}]{muggliatf}%
  \BibitemOpen
  \bibfield  {author} {\bibinfo {author} {\bibfnamefont {P.}~\bibnamefont
  {Muggli}}, \bibinfo {author} {\bibfnamefont {V.}~\bibnamefont {Yakimenko}},
  \bibinfo {author} {\bibfnamefont {M.}~\bibnamefont {Babzien}}, \bibinfo
  {author} {\bibfnamefont {E.}~\bibnamefont {Kallos}}, \ and\ \bibinfo {author}
  {\bibfnamefont {K.~P.}\ \bibnamefont {Kusche}},\ }\href {\doibase
  10.1103/PhysRevLett.101.054801} {\bibfield  {journal} {\bibinfo  {journal}
  {Phys. Rev. Lett.}\ }\textbf {\bibinfo {volume} {101}},\ \bibinfo {pages}
  {054801} (\bibinfo {year} {2008})}\BibitemShut {NoStop}%
\bibitem [{\citenamefont {Emma}\ \emph {et~al.}(2004)\citenamefont {Emma},
  \citenamefont {Bane}, \citenamefont {Cornacchia}, \citenamefont {Huang},
  \citenamefont {Schlarb}, \citenamefont {Stupakov},\ and\ \citenamefont
  {Walz}}]{emmaspoiler}%
  \BibitemOpen
  \bibfield  {author} {\bibinfo {author} {\bibfnamefont {P.}~\bibnamefont
  {Emma}}, \bibinfo {author} {\bibfnamefont {K.}~\bibnamefont {Bane}}, \bibinfo
  {author} {\bibfnamefont {M.}~\bibnamefont {Cornacchia}}, \bibinfo {author}
  {\bibfnamefont {Z.}~\bibnamefont {Huang}}, \bibinfo {author} {\bibfnamefont
  {H.}~\bibnamefont {Schlarb}}, \bibinfo {author} {\bibfnamefont
  {G.}~\bibnamefont {Stupakov}}, \ and\ \bibinfo {author} {\bibfnamefont
  {D.}~\bibnamefont {Walz}},\ }\href {\doibase 10.1103/PhysRevLett.92.074801}
  {\bibfield  {journal} {\bibinfo  {journal} {Phys. Rev. Lett.}\ }\textbf
  {\bibinfo {volume} {92}},\ \bibinfo {pages} {074801} (\bibinfo {year}
  {2004})}\BibitemShut {NoStop}%
\bibitem [{\citenamefont {Emma}\ \emph {et~al.}(2012)\citenamefont {Emma},
  \citenamefont {Zhou}, \citenamefont {Huang},\ and\ \citenamefont
  {Behrens}}]{emma-skewquad}%
  \BibitemOpen
  \bibfield  {author} {\bibinfo {author} {\bibfnamefont {P.}~\bibnamefont
  {Emma}}, \bibinfo {author} {\bibfnamefont {F.}~\bibnamefont {Zhou}}, \bibinfo
  {author} {\bibfnamefont {Z.}~\bibnamefont {Huang}}, \ and\ \bibinfo {author}
  {\bibfnamefont {C.}~\bibnamefont {Behrens}},\ }\href@noop {} {\bibfield
  {journal} {\bibinfo  {journal} {Proceedings of the Free Electron Laser
  Conference}\ } (\bibinfo {year} {2012})}\BibitemShut {NoStop}%
\bibitem [{\citenamefont {Borland}(2000)}]{elegant}%
  \BibitemOpen
  \bibfield  {author} {\bibinfo {author} {\bibfnamefont {M.}~\bibnamefont
  {Borland}},\ }\href@noop {} {\bibfield  {journal} {\bibinfo  {journal}
  {Advanced Photon Source LS-287}\ } (\bibinfo {year} {2000})}\BibitemShut
  {NoStop}%
\bibitem [{\citenamefont {Brinkmann}\ \emph {et~al.}(2001)\citenamefont
  {Brinkmann}, \citenamefont {Derbenev},\ and\ \citenamefont
  {Fl\"ottmann}}]{brinkmanflat}%
  \BibitemOpen
  \bibfield  {author} {\bibinfo {author} {\bibfnamefont {R.}~\bibnamefont
  {Brinkmann}}, \bibinfo {author} {\bibfnamefont {Y.}~\bibnamefont {Derbenev}},
  \ and\ \bibinfo {author} {\bibfnamefont {K.}~\bibnamefont {Fl\"ottmann}},\
  }\href {\doibase 10.1103/PhysRevSTAB.4.053501} {\bibfield  {journal}
  {\bibinfo  {journal} {Phys. Rev. ST Accel. Beams}\ }\textbf {\bibinfo
  {volume} {4}},\ \bibinfo {pages} {053501} (\bibinfo {year}
  {2001})}\BibitemShut {NoStop}%
\bibitem [{\citenamefont {Piot}\ \emph {et~al.}(2006)\citenamefont {Piot},
  \citenamefont {Sun},\ and\ \citenamefont {Kim}}]{piotflat}%
  \BibitemOpen
  \bibfield  {author} {\bibinfo {author} {\bibfnamefont {P.}~\bibnamefont
  {Piot}}, \bibinfo {author} {\bibfnamefont {Y.-E.}\ \bibnamefont {Sun}}, \
  and\ \bibinfo {author} {\bibfnamefont {K.-J.}\ \bibnamefont {Kim}},\ }\href
  {\doibase 10.1103/PhysRevSTAB.9.031001} {\bibfield  {journal} {\bibinfo
  {journal} {Phys. Rev. ST Accel. Beams}\ }\textbf {\bibinfo {volume} {9}},\
  \bibinfo {pages} {031001} (\bibinfo {year} {2006})}\BibitemShut {NoStop}%
\bibitem [{\citenamefont {Piot}\ \emph
  {et~al.}(2013{\natexlab{a}})\citenamefont {Piot}, \citenamefont {Prokop},
  \citenamefont {Carlsten}, \citenamefont {Mihalcea},\ and\ \citenamefont
  {Sun}}]{astaflat}%
  \BibitemOpen
  \bibfield  {author} {\bibinfo {author} {\bibfnamefont {P.}~\bibnamefont
  {Piot}}, \bibinfo {author} {\bibfnamefont {C.}~\bibnamefont {Prokop}},
  \bibinfo {author} {\bibfnamefont {B.}~\bibnamefont {Carlsten}}, \bibinfo
  {author} {\bibfnamefont {D.}~\bibnamefont {Mihalcea}}, \ and\ \bibinfo
  {author} {\bibfnamefont {Y.}~\bibnamefont {Sun}},\ }\href@noop {} {\bibfield
  {journal} {\bibinfo  {journal} {Proceedings of the International Particle
  Accelerator Conference}\ } (\bibinfo {year}
  {2013}{\natexlab{a}})}\BibitemShut {NoStop}%
\bibitem [{\citenamefont {Piot}\ \emph
  {et~al.}(2013{\natexlab{b}})\citenamefont {Piot}, \citenamefont {Shiltsev},
  \citenamefont {Nagaitsev}, \citenamefont {Church}, \citenamefont {Garbincius}
  \emph {et~al.}}]{asta}%
  \BibitemOpen
  \bibfield  {author} {\bibinfo {author} {\bibfnamefont {P.}~\bibnamefont
  {Piot}}, \bibinfo {author} {\bibfnamefont {V.}~\bibnamefont {Shiltsev}},
  \bibinfo {author} {\bibfnamefont {S.}~\bibnamefont {Nagaitsev}}, \bibinfo
  {author} {\bibfnamefont {M.}~\bibnamefont {Church}}, \bibinfo {author}
  {\bibfnamefont {P.}~\bibnamefont {Garbincius}},  \emph {et~al.},\ }\href@noop
  {} {\  (\bibinfo {year} {2013}{\natexlab{b}})},\ \Eprint
  {http://arxiv.org/abs/1304.0311} {arXiv:1304.0311 [physics.acc-ph]}
  \BibitemShut {NoStop}%
%%CITATION = ARXIV:1304.0311;%%
\bibitem [{\citenamefont {Nasse}\ \emph {et~al.}(2013)\citenamefont {Nasse},
  \citenamefont {Schuh}, \citenamefont {Naknaimueang}, \citenamefont {Schwarz},
  \citenamefont {Plech}, \citenamefont {Mathis}, \citenamefont {Rossmanith},
  \citenamefont {Wesolowski}, \citenamefont {Huttel}, \citenamefont
  {Schmelling},\ and\ \citenamefont {Muller}}]{flute}%
  \BibitemOpen
  \bibfield  {author} {\bibinfo {author} {\bibfnamefont {M.~J.}\ \bibnamefont
  {Nasse}}, \bibinfo {author} {\bibfnamefont {M.}~\bibnamefont {Schuh}},
  \bibinfo {author} {\bibfnamefont {S.}~\bibnamefont {Naknaimueang}}, \bibinfo
  {author} {\bibfnamefont {M.}~\bibnamefont {Schwarz}}, \bibinfo {author}
  {\bibfnamefont {A.}~\bibnamefont {Plech}}, \bibinfo {author} {\bibfnamefont
  {Y.-L.}\ \bibnamefont {Mathis}}, \bibinfo {author} {\bibfnamefont
  {R.}~\bibnamefont {Rossmanith}}, \bibinfo {author} {\bibfnamefont
  {P.}~\bibnamefont {Wesolowski}}, \bibinfo {author} {\bibfnamefont
  {E.}~\bibnamefont {Huttel}}, \bibinfo {author} {\bibfnamefont
  {M.}~\bibnamefont {Schmelling}}, \ and\ \bibinfo {author} {\bibfnamefont
  {A.-S.}\ \bibnamefont {Muller}},\ }\href {\doibase 10.1063/1.4790431}
  {\bibfield  {journal} {\bibinfo  {journal} {Review of Scientific
  Instruments}\ }\textbf {\bibinfo {volume} {84}},\ \bibinfo {eid} {022705}
  (\bibinfo {year} {2013})}\BibitemShut {NoStop}%
\end{thebibliography}%

\end{document}